\documentclass[aps,prd,twocolumn,superscriptaddress,nofootinbib,floatfix]{revtex4}
\usepackage[dvips]{graphicx}% Include figure files
\usepackage{amsmath,amssymb,latexsym}
\usepackage{bm}% bold math
\usepackage{epsfig}

\newcommand{\nutau}{\nu_\tau}
\newcommand{\nue}{\nu_e}
\newcommand{\numu}{\nu_\mu}
\newcommand{\Enu}{E_{\nu}}

\newcommand{\Nnue}{N_{\nu_e}}
\newcommand{\Nnumu}{N_{\nu_\mu}}

\newcommand{\Epho}{E_\gamma}
\newcommand{\Epi}{E_\pi}

\newcommand{\Npho}{N_\gamma}

\newcommand{\postscript}[2]{\setlength{\epsfxsize}{#2\hsize}
   \centerline{\epsfbox{#1}}}

\begin{document}

\title{Neutrino Flux from Cosmic Ray Accelerators in the Cygnus Spiral Arm of the Galaxy}

\author{Luis Anchordoqui}
\affiliation{Department of Physics,
University of Wisconsin-Milwaukee, P.O. Box 413, Milwaukee, WI 53201, USA
%\PRE{\vspace*{.1in}}
}

\author{Francis Halzen}
\affiliation{Department of Physics,
University of Wisconsin, Madison, WI 53706, USA}

\author{Teresa Montaruli}
\affiliation{Department of Physics,
University of Wisconsin, Madison, WI 53706, USA}
\affiliation{Universit\'a di Bari, Via Amendola 173, 70126 Bari, Italy}

\author{Aongus O'Murchadha}
\affiliation{Department of Physics,
University of Wisconsin, Madison, WI 53706, USA}

\date{December 2006}
\begin{abstract}
  \noindent Intriguing evidence has been accumulating for the
  production of cosmic rays in the Cygnus region of the Galactic
  plane. We here show that the IceCube experiment can produce
  incontrovertible evidence for cosmic ray acceleration by observing
  the neutrinos from the decay of charged pions accompanying the TeV
  photon flux observed in the HEGRA, Whipple, Tibet and Milagro
  experiments.  Our assumption is that the TeV photons observed are
  the decay products of neutral pions produced by cosmic ray
  accelerators in the nearby spiral arm of the galaxy. Because of the
  proximity of the sources, IceCube will obtain evidence at the
  $5\sigma$ level in 15 years of observation.

\end{abstract}

%\pacs{xxxx}

\maketitle

Evidence may be emerging for a cosmic accelerator in the Cygnus spiral
arm. The observation of the Cygnus region by the HEGRA IACT-system has
allowed the serendipitous discovery of a TeV $\gamma$-ray 
source~\cite{Aharonian:2002ij}, with an average flux $\sim 3\%$ of the
Crab Nebula~\cite{Crab}. The analysis of the total 278.3 hours of
observations performed in two periods from 1999 to 2002 (120.5 hours
from 1999 to 2001~\cite{Aharonian:2002ij} and 157.8 hours during
2002~\cite{Aharonian:2005ex}) has revealed the presence of a steady
(and possibly extended) TeV source, with hard injection spectrum.

The excess significance of the TeV source is $7.1\sigma$ and it
appears extended at more than $4\sigma$ level with a morphology which
is suitably described by a Gaussian profile. The source is termed TeV
J2032+4130 after the position of the center of gravity and its
extension (Gaussian $1\sigma$ radius) is $6.2'(\pm1.2'_{\rm stat} \pm
0.9'_{\rm sys})$. Especially intriguing is the possible association of
TeV J2032+4130 with Cygnus OB2, a cluster of more than 2700
(identified) young, hot stars with a total mass of $\sim 10^4$ solar
masses~\cite{Knodlseder:2000vq}.  At a relatively small distance
($\approx 5000$~light years) to Earth~\cite{note2}, this is the
largest massive Galactic stellar association.

The observed hot spot has no clear counterpart and the spectrum is not
easily accommodated with synchrotron radiation by electrons. The
difficulty to accommodate the spectrum by conventional electromagnetic
mechanisms has been exacerbated by the failure of CHANDRA and VLA to
detect X-rays or radiowaves signaling acceleration of any
electrons~\cite{Butt:2003xc}.  The two most plausible models to
explain the $\gamma$-ray signal are: $(a)$ A proton beam (accelerated
in the stellar winds of Cygnus
OB2~\cite{Aharonian:2002ij,Butt:2003xc}, or else in the wind nebulae
of an undetected nearby pulsar~\cite{Bednarek:2003cx})
interacting with a molecular cloud to produce pions that are the
source of the gamma rays. Proton acceleration to explain the TeV
photon signal requires only 1\% efficiency for the conversion of the
energy in the stellar wind into cosmic ray acceleration. $(b)$ The TeV
gamma rays can also originate in the photo-deexcitation of
ultra-relativistic nuclei (Lorentz factor $\approx 10^{6}$) that are
themselves the photo-disintegration products of heavier nuclei
broken-up in the bath of intense UV photons from the Lyman $\alpha$
emissions of hot stars~\cite{Anchordoqui:2006pd}. As in the proton
beam model, the required power density for acceleration of nuclei is 2
orders of magnitude smaller than the kinetic energy budget of the
entire association.

An additional set of observations performed during 1989-1990 by the
Whipple Observatory~\cite{Lang:2004bk} has been recently reanalyzed in
the light of the HEGRA data. These confirm an excess in the same
direction as J2032+4130, although with considerably larger average
flux ($\sim 12\%$ of the Crab), above a peak energy response of
0.6~TeV.  The statistical significance of the signal is only 10\%
smaller with selection of events above 1.2~TeV. However, the large
differences between the flux levels cannot be explained as errors in
estimation of the sensitivity of the experiments since they have been
calibrated by the simultaneous observations of other TeV sources. More
recently, data taken with the Whipple Observatory during 2003-2005 have 
been reported~\cite{Konopelko:2006jr}.  The analysis of the latest
dataset reveals a TeV hot spot (integral flux $\sim 8\%$ of the Crab)
that is displaced about 9 arcminutes to the northeast of the TeV
J2032+4130 position. A re-analysis~\cite{Butt:2006js} of a 10-hour VLA 
mosaic exposure towards 
TeV J2032+4130 (based on the alternative source hypothesis)
allowed the detection of a weak, predominantly non-thermal, 
shell-like supernova remnant-type object
(with location and morphology very similar to the HEGRA source)
that can be the cosmic ray engine powering the OB association.

Very recently, the Milagro Collaboration reported an excess of events
from the Cygnus region at the $10.9\sigma$ level~\cite{Abdo:2006fq}.
The observed flux within a $3^\circ \times 3^\circ$ 
window centered at the HEGRA source is 70\% of the Crab at the median
detected energy of 12~TeV, and has a differential spectrum $\propto
E^{-2.6}.$ Such a flux largely exceeds the one reported by the HEGRA
Collaboration, implying that there could be a population of 
unresolved TeV $\gamma$-ray sources within the Cygnus OB2 association.

The Milagro Collaboration also reported a new hot spot, christened
MGRO J2019+37, at right assension $= 304.83^\circ \pm 0.14_{\rm stat}
\pm 0.3_{\rm sys}$ and declination $= 36.83^\circ \pm 0.08_{\rm stat}
\pm 0.25_{\rm sys}$~\cite{Abdo:2006fq}. This new unidentified
source is observed with statistical significance $>6\sigma$ above the
average diffuse $\gamma$-ray emission in the region. A fit to a circular
2-dimensional Gaussian yields a width of $0.32 \pm 0.12$ degrees,
which for a distance of 1.7~kpc suggests a source radius of about
9~pc. For a differential spectrum $\propto E^{-2.6}$, the brightest
hotspot in the Milagro map of the Cygnus region represents a flux of 1
Crab above 12.5~TeV. Interestingly, the Tibet AS-gamma Collaboration
has observed a cosmic ray anisotropy from the direction of Cygnus,
which is consistent with Milagro's
measurements~\cite{Amenomori:2006bx}. Unfortunately, the Tibet array
has very little power to distinguish how much of the anisotropy should
be attributed to $\gamma$-rays and how much, if any, to baryons.

The brightest Milagro hot spot is located outside the OB association.
However, the $\gamma$-ray signal is found to trace the gas density
distribution in the region. The model proposed is that of a cosmic ray
beam, perhaps powered by a millisecond pulsar, which interacts with a
molecular cloud positioned a few degrees to the southeast of the OB
star cluster. If the $\gamma$-ray emission from MGRO J2019+37
originates in $\pi^0$ decay, it is necessarily accompanied by a flux
of high energy neutrinos emerging from the $\pi^\pm$ population. In
this paper we discuss in detail the prospects to observe such a flux
with the IceCube neutrino telescope.

The pion spectrum resulting from collisions of the ultrarelativistic
protons on the molecular cloud is expected to obey a modified
Feynman scaling in the central rapidity region, $dN_\pi/dE_\pi|_{E_p}
\approx C(E_p)/E_\pi,$ where $C$ may be growing as some power of $\ln
E_p$~\cite{Abe:1989td}. For given $E_\pi\ < 0.08 \,E_{p,{\rm max}},$
we may convolve with a proton spectrum typical of Fermi engines, $
dN_p/dE_p \propto E_p^{-\Gamma},$ to obtain the pion
spectrum~\cite{Anchordoqui:2004bd}
\begin{equation}
\frac{dN_\pi}{dE_\pi}  =  \int_{E_\pi/0.08}^{E_{p, {\rm max}}}
dE_p\; \left. \frac{dN_\pi}{dE_\pi}
\right|_{E_p}\frac{dN_p}{dE_p}\;\; \propto\;\; \frac{ \bar
C(E_\pi)}{E_\pi^{\Gamma}}\,\, , \label{soft1}
\end{equation}
where $\bar C(E_\pi)$ is generically a function which grows as a power
of $\ln E_\pi$, falling to zero at the cutoff $E_\pi=0.08\ E_{p,{\rm
    max}}$. Since $\pi^0$'s, $\pi^+$'s, and $\pi^-$'s are made in
equal numbers, one expects two photons, two $\nue$'s, and four
$\numu$'s per $\pi^0$. Gamma rays, produced via $\pi^0$ decay carry
one-half of the energy of the pion. Each $\pi^-$ decays to 3 neutrinos
and an electron, $\pi^- \to \mu^- \overline\nu_\mu \to \nu_\mu
\overline \nu_\mu \overline \nu_e e^-$.  The electron radiatively cools
through interactions with the gas and the ambient magnetic and
radiation fields. Typically $e^-$ synchrotron emission extends from
radio frequencies to X-rays. The average neutrino energy from the
direct pion decay is $\langle E_{\nu_\mu} \rangle_\pi = (1-r)\,E_\pi/2
\simeq 0.22\,E_\pi$ and that of the muon is $\langle E_{\mu}
\rangle_\pi = (1+r)\,E_\pi/2 \simeq 0.78\,E_\pi$, where $r$ is the
ratio of muon to the pion mass squared. Now, taking the $\nu_\mu$ from
muon decay to have 1/3 the energy of the muon, the average energy of
the $\nu_\mu$ from muon decay is $\langle E_{\nu_\mu} \rangle_\mu =
(1+r)E_\pi/6=0.26 \, E_\pi$.  Similar considerations apply for the
charged conjugate process.  For simplicity, hereafter we consider that
all neutrinos carry one quarter of the energy of the pion. The
energy-bins $dE$ scale with these fractions, and we arrive 
at~\cite{Anchordoqui:2004eu}
\begin{eqnarray}
\frac{d\Npho}{d\Epho} (\Epho=\Epi/2) & = &
    4\,\frac{dN_{\pi}}{d\Epi}(\Epi)\,, \nonumber \\
\frac{d\Nnue}{d\Enu} (\Enu= \Epi/4 ) & = &
    8\,\frac{dN_{\pi}}{d\Epi}(\Epi)\,, \\
\frac{d\Nnumu}{d\Enu } (\Enu = \Epi/4 ) & = &
    16\,\frac{dN_{\pi}}{d\Epi}(\Epi)\,,\nonumber
\label{ebin}
\end{eqnarray}
where $\pi$ denotes any one of the three pion charge-states.

Whereas the details are complex and predictions can be treacherous, it
is clear that the astrophysical ambiguities far outweigh the details
associated with the particle physics, and hence it is safe to assume
that identical fluxes of $\gamma$-rays and $\nu_\mu$ are produced.
Terrestrial experiments have shown that $\numu$ and $\nutau$ are
maximally mixed with a mass-squared difference $\sim 10^{-3}{\rm
  eV}^2$, and that $|\langle \nu_e|\nu_3\rangle|^2$ is nearly
zero~\cite{Gonzalez-Garcia:2004jd}.  Here $\nu_3 \simeq
(\nu_\mu+\nu_\tau)/\sqrt{2}$ is the third neutrino eigenstate. This
implies that any initial flavor ratio having $\omega_e = 1/3$ will
arrive at Earth with ratios $\omega_e : \omega_\mu : \omega_\tau =
1:1:1.$ Thus, there is a fairly robust prediction that the initial
flavor ratios of $1:2:0$ given in Eq.~(\ref{ebin}) would arrive at
Earth democratically distributed, i.e., $1:1:1.$ From these remarks,
one finds a nearly identical flux,
\begin{eqnarray}
\frac{dF_{\nu_\alpha}}{dE_{\nu}} & = & \frac{1}{4 \pi d^2} 
  \,\,\frac{d\dot N_{\nu_e}}{dE_\nu} \nonumber \\
  & \approx & 7.7 \times 10^{-12} \left(\frac{E_\nu}{\rm TeV}\right)^{-2.6} {\rm TeV}^{-1} {\rm cm}^{-2} {\rm s}^{-1}
\label{fnu}
\end{eqnarray}
for each of the three neutrino flavors $\alpha=e,\mu,\tau$~\cite{Alvarez-Muniz:2002tn}.

The Antarctic Muon And Neutrino Detector Array
(AMANDA)~\cite{Andres:1999hm}, using natural 1 mile deep Antarctic ice
as a \v {C}erenkov detector, has operated for more than 5 years in its
final configuration: 19 strings instrumented with 680 optical modules.
IceCube~\cite{Ahrens:2003ix}, the successor experiment to AMANDA, is
now under construction. It will consist of 80 kilometer-length
strings, each instrumented with 60 digital optical modules (DOM)
spaced by 17~m.  The deepest module is 2.4~km below the surface. The
strings are arranged at the apexes of equilateral triangles 125\,m on
a side. The instrumented (not effective!)  detector volume is a full
cubic kilometer. A surface air shower detector, IceTop, consisting of
160 \v {C}erenkov detectors deployed over 1\,km$^{2}$ above IceCube,
augments the deep-ice component by providing a tool for calibration,
background rejection and air-shower physics. The angular resolution
for muon tracks $\approx 0.7^\circ$~\cite{Ahrens:2002dv} allows a
search window of $1^\circ \times 1^\circ$. Construction of the
detector started in the Austral summer of 2004/2005 and will continue
for 6 years, possibly less. At the time of writing, data collection by
the first 22 strings and 52 IceTop stations has begun.

\begin{figure}
 \postscript{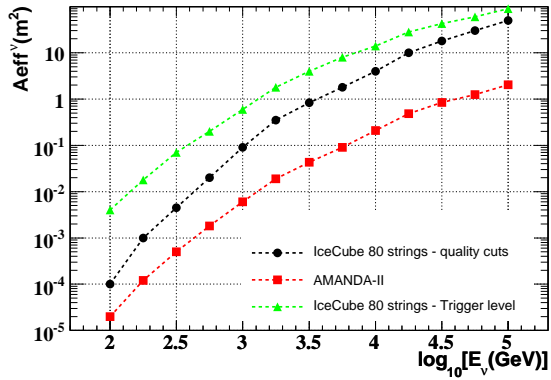}{0.98}
 \caption{The IceCube effective area for muon track reconstruction as
   a function of the neutrino energy. Two curves are shown indicating
   the trigger level and the recent estimate with very conservative
   quality cuts. For comparison the effective area of AMANDA II is
   also shown.}
\label{fig:1}
\end{figure}

The event rate from a source located at given declination can be
calculated from the knowledge of the so-called neutrino effective
area.  This parameter strongly depends on the energy due to the almost
linear increase of the neutrino interaction cross-section and to the
logaritmic rise of the muon range at very high energies. Moreover it
accounts for the probability of absorption of neutrinos during their
propagation through the Earth, where they may disappear due to a
charged current interaction. It should also account for the detection
and reconstruction efficiency. Preliminary studies by the IceCube
Collaboration~\cite{Achterberg:2006pw}, using data collected with the
9 In-Ice strings and the 16 IceTop stations, show a good agreement
with the detector performance~\cite{Ahrens:2003ix}. The IceCube
simulation software is currently under active development and
extrapolations of the 9-string results are undeway. In particular, a
conservative estimate of the effective area for reconstruction of
muons tracks, with severe cuts to obtain a good angular resolution and
rejection of misreconstructed muons has been estimated
in~\cite{Achterberg:2006pw}.  Such an effective area is shown in
Fig.~\ref{fig:1} together with the trigger level effective area
(defined as 8 DOM threshold) of IceCube and the effective area of the
AMANDA final configuration.

\begin{figure}
 \postscript{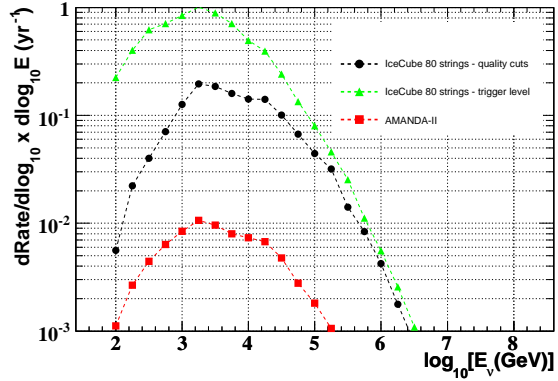}{0.98}
 \caption{Differential energy distribution of event rates from MGRO 
J2019+37. The different curves indicate rates computed using the
   effective area $(a)$ of the trigger level condition, $(b)$ of the
   conservative quality cuts, $(c)$ of the AMANDA II array.}
\label{fig:2}
\end{figure}

Equipped with the effective area shown in Fig.~\ref{fig:1} it is
straightforward to calculate lower and upper limits on the muon
neutrino event rate from the Cygnus region
\begin{equation}
{\rm Rate} \equiv \left. 
\frac{d{\cal N}_{\rm S}}{dt} \right|_{\delta = 36.8^\circ} = 
\int A_{\rm eff}^{\nu}(E_{\nu}) \,\, \frac{dF_{\nu_\mu}}{
dE_{\nu}}\,\, dE_{\nu} \, .
\label{rate}
\end{equation}
Using the neutrino flux given in Eq.~(\ref{fnu}) and convolving it
with the IceCube neutrino effective area we foresee an event rate of
muon neutrinos with $E_\nu > 1~{\rm TeV}$ of $1.1~{\rm yr}^{-1} < d {\cal N}_{\rm S}/dt \alt 4.1~{\rm
  yr}^{-1}.$ The energy distribution of such events is shown in
Fig.~\ref{fig:2}.

We now turn to the estimate of the background.  For the atmospheric
neutrino flux, arising from the decay of pions and kaons produced in
cosmic ray interactions with the air molecules, we adopt the estimates
of Ref.~\cite{Barr:2004br}.  We obtain the number of expected muon
tracks from atmospheric neutrinos as in Eq.~(\ref{rate}), using the
$\nu_\mu$ atmospheric neutrino flux integrated over a solid angle of
$1^\circ\times 1^\circ$ width around the direction of the MGRO
J2019+37 (zenith angle $\theta=53.2^\circ$). We obtain an expected
background of atmospheric tracks, $1.2~{\rm yr}^{-1} < d{\cal N}_{\rm
  B}/dt \alt 5.5~{\rm yr}^{-1}.$

These event rates are based on a conservative estimate of the level of
detail to which we currently understand the detector performance.
However, since this understanding will improve over time one expects
the systematic errors to decrease to the levels projected in the
baseline design reported in~\cite{Ahrens:2003ix}. Therefore, to
determine the discovery reach we employ the semianalytical calculation
presented in~\cite{Gonzalez-Garcia:2005xw} based on a full Monte Carlo
simulation using these projected baseline detector properties, with
quality cuts referred as level 2 cuts~\cite{note}. For a muon energy
threshold of 100~GeV and minimum track length of 300~m, the expected
rate of $\nu_\mu$ induced tracks is $d {\cal N}_{\rm S}/dt \simeq
3~{\rm yr}^{-1}$, with a background of $d {\cal N}_{\rm B}/dt \simeq
2.5~{\rm yr}^{-1}.$ Hence, after 15~yr of operation, the (total) 
detection significance,
\begin{equation}
S_{\rm det} = \frac{{\cal N}_{\rm S}}{ \sqrt{{\cal N}_{\rm B} 
+ {\cal N}_{\rm S}}}  
\simeq 5\sigma \,\,, 
\end{equation}
is expected to be at discovery level.

We now verify that our results are consistent with existing data. Very
recently, the AMANDA II data collected during 2000 - 2004 (with a
lifetime of 1001 days) was analyzed to set new limits on the neutrino
fluxes from point sources (circular bin size varying
between $2.25^\circ$ and $3.75^\circ$ depending on
declination)~\cite{Achterberg:2006vc}. For the effective
area shown in Fig.~\ref{fig:1}, the expected background from
atmospheric neutrinos $(E_\nu > 100~{\rm GeV})$ in a 
$1^\circ \times 1^\circ$ window is $d {\cal
  N}_{\rm B}/dt \simeq 0.22~{\rm yr}^{-1}$. The AMANDA Collaboration
reported an expected atmospheric neutrino background from the
direction of J2032+4130 of 6.8 events, in complete agreement with 
our calculations. The experiment observed 7 events pointing towards the same 
direction of the sky, leading to a 90\% CL upper limit,
\begin{equation}
E^2_\nu \frac{dF_{\nu_\alpha}}{dE_{\nu}} =  1.1 \times 10^{-10}~{\rm TeV} \, 
{\rm cm}^{-2}\, {\rm s}^{-1} \,.
\end{equation}
Thus, the sensitivity reach at AMANDA cannot probe the predicted flux 
given in Eq.~(\ref{fnu}).

In summary, by observing the neutrinos from the decay of charged pions
accompanying the recently detected $\gamma$-rays with the Milagro
experiment, the IceCube neutrino telescope will produce
incontrovertible evidence for cosmic ray acceleration in the Cygnus
spiral arm. In this paper, we have discussed in detail the sensitivity
reach of IceCube to the brightest hot spot, MGRO J2019+37.
Contributions from TeV $\gamma$-ray hot spots (still hidden) within
the Cygnus OB2 association will certainly enhance the
signal~\cite{Anchordoqui:2003vc}. Moreover, as suggested recently by
the Milagro Collaboration~\cite{Abdo:2006fq}, to smoothly match EGRET
data in the 100 MeV energy region, the spectra from all the sources in
the Cygnus region should have a break (perhaps because of absorption
effects) and be harder than $E^{-2.6}$ at lower energies. As an
illustration, we have estimated the expected event rate from MGRO
J2019+37, using the semianalytical calculation given
in~\cite{Gonzalez-Garcia:2005xw} and assuming a spectrum $\propto
E_\nu^{-2.4},$
\begin{equation} 
d {\cal N}_{\rm S}/dt \simeq 9.0~{\rm yr}^{-1} \, .
\end{equation}
For a background of $2.5~{\rm yr}^{-1}$ events, this implies that IceCube will attain a $5 \sigma$ discovery reach in 2 years of operation!

\acknowledgments{We would like to thank Juande Zornoza for deriving
  the effective areas shown in Fig.~\ref{fig:1}. This work has been
  supported in part by the US NSF under Grant No. OPP- 0236449, in
  part by the US Department of Energy (DoE) Grant No.
  DE-FG02-95ER40896, in part by the University of Wisconsin Research
  Committee with funds granted by the Wisconsin Alumni Research
  Foundation, and in part by the University of Wisconsin-Milwaukee.}


\begin{thebibliography}{99}


\bibitem{Aharonian:2002ij}
  F.~A.~Aharonian, A.~Akhperjanian, M.~Beilicke, Y.~Uchiyama and T.~Takahashi,
  %``An unidentified TeV source in the vicinity of Cygnus OB2,''
  Astron.\ Astrophys.\  {\bf 393}, L37 (2002).
  %[arXiv:astro-ph/0207528].
  %%CITATION = ASTRO-PH 0207528;%%


\bibitem{Crab} The integral $\gamma$-ray flux obtained from the Crab
  by the Whipple Collaboration is now the standard TeV~flux unit:
  $F_{\rm Crab}(E_\gamma>0.35\,{\rm TeV})= 10^{-10}~{\rm cm}^2\, {\rm
    s}^{-1}$.  The spectral index of the Crab's integrated flux is
measured to be $-1.5$. G.~Vacanti {\em et al.},
Astrophys. J. 377, 467 (1991); A.~M.~Hillas {\em et al.}, Astrophys.
J. 503, 744 (1998).





\bibitem{Aharonian:2005ex}
  F.~Aharonian {\it et al.}  [The HEGRA Collaboration],
  %``The unidentified TeV source (TeV J2032+4130) and surrounding field:  Final
  %HEGRA IACT-system results,''
  Astron.\ Astrophys.\  {\bf 431}, 197 (2005).
  %[arXiv:astro-ph/0501667].
  %%CITATION = ASTRO-PH 0501667;%%  




\bibitem{Knodlseder:2000vq}
  J.~Knodlseder,
  %``Cygnus OB2 - a young globular cluster in the Milky Way,''
  arXiv:astro-ph/0007442.
  %%CITATION = ASTRO-PH 0007442;%%

\bibitem{note2} For such a distance, the TeV HEGRA source would have a 
radius $ \sim 3.07 (\pm 0.59_{\rm stat} \pm 0.45_{\rm sys})~{\rm pc}$.


\bibitem{Butt:2003xc}
  Y.~Butt {\it et al.},
  %``CHANDRA/VLA follow-up of TeV J2032+4131, the only unidentified TeV
  %gamma-ray source,''
  Astrophys.\ J.\  {\bf 597}, 494 (2003);
  %[arXiv:astro-ph/0302342];
  %%CITATION = ASTRO-PH 0302342;%%
%\bibitem{Butt:2005dx}
  Y.~Butt, J.~Drake, P.~Benaglia, J.~Combi, T.~Dame, F.~Miniati and G.~Romero,
  %``Deeper Chandra Follow-up of Cygnus TeV Source Perpetuates Mystery,''
  Astrophys.\ J.\  {\bf 643}, 238 (2006).
  %[arXiv:astro-ph/0509191].
  %%CITATION = ASTRO-PH 0509191;%%


\bibitem{Bednarek:2003cx}
  W.~Bednarek,
  %``Gamma-rays and cosmic-rays from a pulsar in Cygnus OB2,''
  Mon.\ Not.\ Roy.\ Astron.\ Soc.\  {\bf 345}, 847 (2003).
  %[arXiv:astro-ph/0307216].
  %%CITATION = ASTRO-PH 0307216;%%


\bibitem{Anchordoqui:2006pd}
  L.~A.~Anchordoqui, J.~F.~Beacom, H.~Goldberg, S.~Palomares-Ruiz and 
  T.~J.~Weiler,
  %``TeV gamma-rays from photo-disintegration / de-excitation of cosmic-ray
  %nuclei,''
  Phys.\ Rev.\ Lett.\  {\bf 98}, 121101 (2007);
  %[arXiv:astro-ph/0611580].
  %%CITATION = PRLTA,98,121101;%%
%\bibitem{Anchordoqui:2006pe}
%  L.~A.~Anchordoqui, J.~F.~Beacom, H.~Goldberg, S.~Palomares-Ruiz 
%and T.~J.~Weiler,
  %``TeV gamma-rays and neutrinos from photo-disintegration of nuclei in Cygnus
  %OB2,''
  Phys.\ Rev.\  D {\bf 75}, 063001 (2007).
%  [arXiv:astro-ph/0611581].
  %%CITATION = PHRVA,D75,063001;%%


  



\bibitem{Lang:2004bk}
  M.~J.~Lang {\it et al.},
  %``Evidence for TeV gamma ray emission from TeV J2032+4130 in Whipple
  %archival data,''
  Astron.\ Astrophys.\  {\bf 423}, 415 (2004).
  %[arXiv:astro-ph/0405513].
  %%CITATION = ASTRO-PH 0405513;%%



\bibitem{Konopelko:2006jr}
  A.~Konopelko {\it et al.},
  %``Observations of the unidentified TeV gamma-ray source TeV J2032+4130 with
  %the Whipple Observatory 10-m telescope,''
  arXiv:astro-ph/0611730.
  %%CITATION = ASTRO-PH 0611730;%%




\bibitem{Butt:2006js}
  Y.~M.~Butt, J.~A.~Combi, J.~Drake, J.~P.~Finley, A.~Konopelko, M.~Lister 
  and J.~Rodriguez,
  %``TeV J2032+4130: A not-so-dark accelerator?,''
  arXiv:astro-ph/0611731.
  %%CITATION = ASTRO-PH 0611731;%%
  See also VLA proposal $\#$AB1075 (PI: Y. M. Butt).





\bibitem{Abdo:2006fq}
  A.~A.~Abdo {\it et al.},
  %``Discovery of TeV gamma-ray emission from the Cygnus region of 
  %the galaxy,''
  Astrophys.\ J.\  {\bf 658}, L33 (2007).
  %[arXiv:astro-ph/0611691].
  %%CITATION = ASJOA,658,L33;%%

  
\bibitem{Amenomori:2006bx}
  M.~Amenomori  [Tibet AS-gamma Collaboration],
  %``Anisotropy and corotation of galactic cosmic rays,''
  Science {\bf 314}, 439 (2006)
  [arXiv:astro-ph/0610671].
  %%CITATION = ASTRO-PH 0610671;%%



\bibitem{Abe:1989td}
  F.~Abe {\it et al.}  [CDF Collaboration],
  %``Pseudorapidity Distributions Of Charged Particles Produced In Anti-P P
  %Interactions At S**(1/2) = 630-Gev And 1800-Gev,''
  Phys.\ Rev.\ D {\bf 41}, 2330 (1990).
  %%CITATION = PHRVA,D41,2330;%%


\bibitem{Anchordoqui:2004bd}
  L.~Anchordoqui, H.~Goldberg and C.~Nunez,
  %``Probing split supersymmetry with cosmic rays,''
  Phys.\ Rev.\ D {\bf 71}, 065014 (2005).
  %[arXiv:hep-ph/0408284].
  %%CITATION = HEP-PH 0408284;%%



\bibitem{Anchordoqui:2004eu}
  L.~A.~Anchordoqui, H.~Goldberg, F.~Halzen and T.~J.~Weiler,
  %``Neutrino bursts from Fanaroff-Riley I radio galaxies,''
  Phys.\ Lett.\ B {\bf 600}, 202 (2004).
  %[arXiv:astro-ph/0404387].
  %%CITATION = ASTRO-PH 0404387;%%


\bibitem{Gonzalez-Garcia:2004jd}
  M.~C.~Gonzalez-Garcia,
  %``Global analysis of neutrino data,''
  arXiv:hep-ph/0410030.
  %%CITATION = HEP-PH 0410030;%%


\bibitem{Alvarez-Muniz:2002tn} We stress that a similar result is
  obtained by normalizing the neutrino flux in a bolometric fashion.
  J.~Alvarez-Muniz and F.~Halzen,
  %``High-energy neutrinos from the cosmic accelerator RX J1713.7-3946,''
  Astrophys.\ J.\  {\bf 576}, L33 (2002).
  %[arXiv:astro-ph/0205408].
  %%CITATION = ASTRO-PH 0205408;%%



\bibitem{Andres:1999hm}
E.~Andres {\it et al.} [AMANDA Collaboration],
%``The AMANDA neutrino telescope: Principle of operation and first  results,''
Astropart.\ Phys.\  {\bf 13}, 1 (2000).
%[arXiv:astro-ph/9906203].
%%CITATION = ASTRO-PH 9906203;%%


\bibitem{Ahrens:2003ix}
J.~Ahrens {\it et al.}  [IceCube Collaboration],
%``Sensitivity of the IceCube detector to astrophysical sources of high
%energy muon neutrinos,''
Astropart.\ Phys.\  {\bf 20}, 507 (2004).
%[arXiv:astro-ph/0305196].
%%CITATION = ASTRO-PH 0305196;%%


\bibitem{Ahrens:2002dv}
J.~Ahrens {\it et al.}  [IceCube Collaboration],
%``IceCube: The next generation neutrino telescope at the South Pole,''
Nucl.\ Phys.\ Proc.\ Suppl.\  {\bf 118}, 388 (2003).
%[arXiv:astro-ph/0209556].
%%CITATION = ASTRO-PH 0209556;%%

\bibitem{Achterberg:2006pw}
  A.~Achterberg  [IceCube Collaboration],
  %``Contributions to 2nd TeV particle astrophysics conference (TeV PA II)
  %Madison Wisconsin - 28-31 August 2006,''
  arXiv:astro-ph/0611597.
  %%CITATION = ASTRO-PH 0611597;%%


\bibitem{Barr:2004br}
G.~D.~Barr, T.~K.~Gaisser, P.~Lipari, S.~Robbins and T.~Stanev,
%``A three-dimensional calculation of atmospheric neutrinos,''
Phys.\ Rev.\ D {\bf 70}, 023006 (2004). 
%[arxiv:astro-ph/0403630].
%%CITATION = ASTRO-PH 0403630;%%




\bibitem{Gonzalez-Garcia:2005xw}
  M.~C.~Gonzalez-Garcia, F.~Halzen and M.~Maltoni,
  %``Physics reach of high-energy and high-statistics Icecube
  %atmospheric neutrino data,''
  Phys.\ Rev.\ D {\bf 71}, 093010 (2005).
  %[arXiv:hep-ph/0502223].
  %%CITATION = HEP-PH 0502223;%%


\bibitem{Achterberg:2006vc}
  A.~Achterberg  [IceCube Collaboration],
  %``Five years of searches for point sources of astrophysical neutrinos 
  %with the AMANDA II neutrino telescope,''
  arXiv:astro-ph/0611063.
  %%CITATION = ASTRO-PH 0611597;%%

\bibitem{note} The effective detector used in our benchmark
  calculations represents a conservative
  prediction of IceCube performance. It is based on a simulation
  using AMANDA technology and analysis techniques. IceCube's digital
  signals, combined with the first experience with data will
  undoubtedly lead to improvements for the threshold and the effective
  area over what has been anticipated on the basis of simulation. In
  this sense our estimates are conservative.







\bibitem{Anchordoqui:2003vc} It is important to stress that if the flux of
  $\gamma$-rays is produced through nuclei de-excitation process, then
  the associated neutrino signal from decay of neutrons produced in
  the course of the photodisintegration would be observable at
  IceCube.  L.~A.~Anchordoqui, H.~Goldberg, F.~Halzen and T.~J.~Weiler,
   %``Galactic point sources of TeV antineutrinos,''
  Phys.\ Lett.\ B {\bf 593}, 42 (2004).
  %[arXiv:astro-ph/0311002].
  %%CITATION = ASTRO-PH 0311002;%%
  See also Ref.~\cite{Anchordoqui:2006pd} for an up-to-date calculation
  of IceCube's sensitivity.





\end{thebibliography}
\end{document}